# On Some Entertaining Applications of the Concept of Set in Computer Science Course

## Krasimir Yordzhev[*], Hristina Kostadinova[**]

[*]Associate Professor Krasimir Yordzhev, Ph.D., Faculty of Mathematics and Natural Sciences, South-West University, Blagoevgrad, Bulgaria, E-mail: yordzhev@swu.bg

[**]Ph.D. student Hristina Kostadinova, Faculty of Mathematics and Natural Sciences, South-West University, Blagoevgrad, Bulgaria, E-mail: kostadinova@swu.bg

*Abstract:* *Some aspects of programming education are examined in this work. It is emphasised, based on the entertainment value, the most appropriate examples are chosen to demonstrate the different language constructions and data structures. Such an example is the demonstrated algorithm for solving the widespread nowadays "Sudoku" puzzle. This is made, because of the connection with the term set and putting it into practice in the programming. Using the so built program there are solved some combinatorial problems, connected to the Sudoku matrices.*

## INTRODUCTION

The present work is contemplated to help the lecturer in programming in his aspiration for giving an appropriate, interesting and entertaining example of the advantage to use the term set in programming. For this purpose the students have to be familiar with the basic definitions of the set theory and to be skillful at the basic operations with sets. It follows the well-known fact that to be a good programmer it is necessary (but not efficient) to be a good mathematician.

A classical example for the use of sets in programming has become the programming realization of the problem for finding prime numbers using the method Sieve of Eratosthenes [2,5,8]. How to construct a faster algorithm, solving the problem for receiving all $n \times n$ binary matrices, which contain exactly $k$ ones in each row and each column, with the help of the set theory and the operations over sets, is shown in [10].

We will examine the entertaining and actual problem, which is very interesting for the students – algorithms for solving Sudoku. It is widespread puzzle nowadays, which presents in the entertaining pages in most of the newspapers and magazines and in entertaining web sites. Sudoku, or Su Doku, is a Japanese word (or phrase) meaning something like Number Place.

On the other side, Sudoku matrices find an interesting combinatorial application, for example in the design theory [7]. The connection between the set of all $m \times m$ permutation matrices (i.e. binary matrices which contain just one 1 in every row and every column) and the set of all $m \times m$ Sudoku matrices, is shown in [3].

As far as the authors of this study know there is not a universal formula for the number of Sudoku matrices $\sigma_n$ with every natural number $n$. We consider it as an open problem in mathematics. When $n = 3$ in [4] is shown that there are exactly

$$\sigma_3 = 6\,670\,903\,752\,021\,072\,936\,960 =$$
$$= 9! \times 72^2 \times 2^7 \times 27\,704\,267\,971 =$$
$$= 2^{20} \times 3^8 \times 5 \times 7 \times 27\,704\,267\,971 \approx$$
$$\approx 6.671 \times 10^{21}$$

in number Sudoku matrices.

**PROBLEM FORMULATION AND ALGORITHM DESCRIPTION**

Let $n$ is a positive integer and let $m = n^2$. Let $S = (s_{ij})$ is a square $m \times m$ matrix, whose elements in this table are integers, belonging to the closed interval $[1, m]$. The matrix $S$ is divided to $n^2$, $n \times n$ square submatrices, which are not intersected and will be called blocks, with the help of $n-1$ horizontal and $n-1$ vertical lines (the matrix $S$, when $n = 3$, is shown on fig. 1).

| $s_{11}$ | $s_{12}$ | $s_{13}$ | $s_{14}$ | $s_{15}$ | $s_{16}$ | $s_{17}$ | $s_{18}$ | $s_{19}$ |
|---|---|---|---|---|---|---|---|---|
| $s_{21}$ | $s_{22}$ | $s_{23}$ | $s_{24}$ | $s_{25}$ | $s_{26}$ | $s_{27}$ | $s_{28}$ | $s_{29}$ |
| $s_{31}$ | $s_{32}$ | $s_{33}$ | $s_{34}$ | $s_{35}$ | $s_{36}$ | $s_{37}$ | $s_{38}$ | $s_{39}$ |
| $s_{41}$ | $s_{42}$ | $s_{43}$ | $s_{44}$ | $s_{45}$ | $s_{46}$ | $s_{47}$ | $s_{48}$ | $s_{49}$ |
| $s_{51}$ | $s_{52}$ | $s_{53}$ | $s_{54}$ | $s_{55}$ | $s_{56}$ | $s_{57}$ | $s_{58}$ | $s_{59}$ |
| $s_{61}$ | $s_{62}$ | $s_{63}$ | $s_{64}$ | $s_{65}$ | $s_{66}$ | $s_{67}$ | $s_{68}$ | $s_{69}$ |
| $s_{71}$ | $s_{72}$ | $s_{73}$ | $s_{74}$ | $s_{75}$ | $s_{76}$ | $s_{77}$ | $s_{78}$ | $s_{79}$ |
| $s_{81}$ | $s_{82}$ | $s_{83}$ | $s_{84}$ | $s_{85}$ | $s_{86}$ | $s_{87}$ | $s_{88}$ | $s_{89}$ |
| $s_{91}$ | $s_{92}$ | $s_{93}$ | $s_{94}$ | $s_{95}$ | $s_{96}$ | $s_{97}$ | $s_{98}$ | $s_{99}$ |

Fig. 1

Let denote by $A_{kl}, 1 \le k, l \le n$ the blocks in the above described matrix $S = (s_{ij})$. Then by definition if $s_{ij} \in A_{kl}$, then

$$(k-1)n < i \le kn$$

and

$$(l-1) < j \le (l-1).$$

Let $s_{ij}$ belong to the block $A_{kl}$ and let $i$ and $j$ are known. Then it is easy to guess that $k$ and $l$ can be calculated with the help of the formulas

$$k = \left[\frac{i-1}{n}\right] + 1$$

and

$$l = \left[\frac{j-1}{n}\right] + 1,$$

where, as usual, we denote by $[x]$ the function: whole part of the real number $x$.

We say that $S = (s_{ij})$, $1 \le i, j \le m = n^2$ is a **Sudoku matrix** if there is just one number of the set $Z_m = \{1, 2, \ldots, m = n^2\}$ in every row, every column and every block.

The puzzle named Sudoku is widespread nowadays. It is given a Sudoku matrix, in which some of the elements are erased. The missing elements will be equal to 0, if we need this. The task in the puzzle is to restore the missing elements of the Sudoku matrix. It is supposed that the authors of the concrete puzzle have chosen the missing elements, so the problem has only one solution. This condition we will miss and will not reckon with it. In this work we will build our programming product to show the number of every possible solution. If the task has no solution this number has to be zero.

The most popular puzzles Sudoku are when $n = 3$, i.e. $m = 9$.

We are going to describe an algorithm for creating a computer program, which finds all solutions (if there are some) of random Sudoku. For this aim we will use the knowledge of the set theory.

We examine the sets $R_i$, $C_j$ and $B_{kl}$, where $1 \le i, j \le m = n^2$, $1 \le k, l \le n$. For every $i = 1, 2, \ldots, m$, the set $R_j$ consists of all missing numbers in the $i$-th row of the matrix. Analogously we define the sets $C_j$, $j = 1, 2, \ldots, m$ correspondingly for the missing numbers in the $j$-th column and $B_{kl}$, $k, l = 1, 2, \ldots, n$ correspondingly for the missing numbers in the blocks $A_{kl}$ of $S$.

When the algorithm starts working it traverses many times all of the elements $s_{ij} \in S$, such that $s_{ij} = 0$, i.e. these are the elements, which real values we have to find.

Let $s_{ij} = 0$ and let $s_{ij} \in A_{kl}$. We assume

$$P = R_i \cap C_j \cap B_{kl}.$$

Then the following three cases are possible:

i) $P = \phi$ (empty set). The task has no solution in this case;

ii) $P = \{d\}$, $d \in Z_m = \{1, 2, \ldots, m\}$, i.e. $|P|$: the number of the elements of $P$ is equal to 1 ($P$ is a set containing one element). Then the only one possibility for $s_{ij}$ is $s_{ij} = d$, i.e. we have found the unknown value of $s_{ij}$ in this case. After this we remove the common element $d$ from the sets $R_i$, $C_j$ and $B_{kl}$, and then we continue to the next zero element of the matrix $S$ (if there is such an element);

iii) $|P| \ge 2$. Then we can not say anything about the unknown value of $s_{ij}$ and we move on the next missing (zero) element of the matrix $S$.

We traverse all zero elements of the matrix $S$ until one the following events occur:

e1) For some $i, j \in \{1, 2, \ldots, m\}$ is true $s_{ij} = 0$, but $P = R_i \cap C_j \cap B_{kl} = \phi$;

e2) All elements in $S$ become positive;

e3) All zero elements of $S$ are traversed, but it does not occur neither event e1, nor event e2. In other words for all the remaining zero elements in $S$, the above described case iii is always true.

In case that it occurs one of the events e1 or e2, then the procedure stops its work and visualizes the obtained result.

In case that event e3 occurs, then the algorithm has to continue working by using other methods, for example it has to apply the "trial and error" method. In the concrete case this method consists of the following things:

We choose random $s_{ij} \in S$, such that $s_{ij} = 0$ and let $k = \left[\dfrac{i-1}{n}\right] + 1$, $l = \left[\dfrac{j-1}{n}\right] + 1$. Let $P = R_i \cap C_j \cap B_{kl} = \{d_1, d_2, \ldots, d_t\}$. Then for every $d_r \in P$, $r = 1, 2, \ldots, t$ we assume $s_{ij} = d_r$. Such an assuming we call a *random trial*. We count the number of all random trials, until the solution is found, in the programming realization of the algorithm. After this we solve the problem for finding the unknown elements of the Sudoku matrix, which contain one element less than the previous matrix. It is comfortable to use a recursion here. The base of the recursion, i.e. we go out of the procedure if there occurs event e1 or e2. It is absolutely sure that it will happen (i.e. there will not be an infinite cycle"), because when we do the random trials we reduce the number of the zero elements by 1.

The above described algorithm is realized in the Pascal programming language, because of the two main reasons:

1) Pascal is chosen, of the most educational institutions in the system of the secondary education, as a first programming language. This means that the basic aim of this work is realized – to help the teacher in his preparation in the process of finding interesting application of the teaching material. That is the way the presented material becomes more interesting for the pupils.

2) There are built in instruments used to work with sets in the Pascal programming language.

The above described ideas can be realized in other random programming language, for example C++. But in this case we have to look for additional instruments to work with sets – for example the associative containers set and multiset realized in Standard Template Library (STL) [1,5]. It can be used the template class set of the system of computer algebra "Symbolic C++", programming code is given in details in [9]. Of course can be built another class set, and specific methods of this class can be described, as a training. This is a good exercise, having in mind the fact that the cardinal number of the basic ("universal") set is not very big. For example the "standard" puzzle Sudoku has basic set the set of the integers from 1 to 9 plus the empty set.

**SOME COMBINATORIAL APPLICATION**

Obviously, if for solving given Sudoku, there is no need to use the trial and error method, i.e. the problem can be solved by using set theoretical operations (0 random trials), then this Sudoku has a unique solution. Such a Sudoku is shown on figure 2. The opposite proposition is not true, as it is in the next example, shown on figure 3, where there is a unique solution and program has made 218 random trials only to receive this solution, as the number of the random trials (after the final solution is received, the program continues working if it is necessary) is 332.

|   |   |   |   |   |   |   |   | 3 |
|---|---|---|---|---|---|---|---|---|
| 5 |   | 3 |   | 8 | 2 |   | 4 | 1 |
| 6 | 2 | 4 | 5 |   |   |   | 7 |   |
| 7 | 8 | 5 |   | 9 |   | 1 | 6 |   |
| 4 |   | 6 | 2 |   |   | 3 |   |   |
|   | 3 |   | 1 | 7 | 6 |   |   |   |
| 1 |   |   |   |   |   |   |   | 4 |
|   |   | 7 | 8 |   | 2 |   |   |   |
| 3 | 4 | 8 | 6 |   |   | 7 | 1 | 5 |

Fig. 2

| 7 |   |   | 3 |   |   |   |   | 4 |
|---|---|---|---|---|---|---|---|---|
|   | 9 |   |   |   | 1 |   |   |   |
|   | 1 | 2 |   |   |   | 4 |   |   |
|   |   |   | 9 |   | 7 | 4 |   |   |
|   | 5 |   |   | 3 |   |   |   |   |
|   | 8 |   |   |   | 2 |   | 1 |   | 7 |
|   |   | 6 |   |   |   |   |   | 3 |
|   |   |   | 7 |   |   | 5 |   |   |
| 8 | 4 |   |   |   |   |   | 2 | 9 |

Fig. 3

Sometimes the Sudoku authors do not examine if the solution is unique. For example the Sudoku, shown on figure 4 (is published in a newspaper's rubric "Easy Sudoku", the name of the newspaper we will not mention) has 4 different solutions, the program does 16 random trials until the final solution is obtained, and the common number of the random trials is 18.

The Sudoku puzzle shown on figure 5 has no solution, although on a first sight there is not contradiction in the condition. The program has to make 21 random trials to receive the result that this puzzle has no solution.

| | | 4 | 9 | | 3 | 8 | | |
|---|---|---|---|---|---|---|---|---|
| | 2 | | 8 | | | | 9 | 3 |
| 3 | | 8 | | 4 | | | 1 | |
| 7 | | | | 9 | 4 | 1 | | |
| 2 | 4 | | 7 | | | | | 8 |
| | 3 | | 2 | | | | 7 | 5 |
| 8 | | 7 | | 2 | | | 3 | |
| | 5 | | | | 7 | 2 | | 9 |
| | | 2 | | 3 | 8 | | 5 | |

Fig. 4

| | 5 | | 9 | | 2 | | 4 | |
|---|---|---|---|---|---|---|---|---|
| 2 | | 7 | | | | 9 | | 8 |
| | 4 | | 8 | 7 | | 3 | | |
| | 1 | | | 2 | | 8 | | 9 |
| 5 | | 8 | | 9 | | | | 3 |
| 7 | | | 5 | | 3 | | 1 | |
| 3 | | 4 | | 5 | | | | 2 |
| | | | 2 | | 7 | 4 | 8 | |
| | 7 | 2 | | | 8 | | 3 | |

Fig. 5

There are even more combinatorial results with different Sudoku matrices, received by the help of the above described algorithm and some experiments.

- There are 283 576 Sudoku matrices of the kind shown on figure 6;
- There are 6 280 Sudoku matrices of the kind shown on figure 7;
- There are 680 Sudoku matrices of the kind shown on figure 8;
- There are 1 728 Sudoku matrices of the kind shown on figure 9;
- There are 22 Sudoku matrices of the kind shown on figure 10;
- There are 8 Sudoku matrices of the kind shown on figure 11;

| 1 | 2 | 3 | * | * | * | * | * | * |
|---|---|---|---|---|---|---|---|---|
| 4 | 5 | 6 | * | * | * | * | * | * |
| 7 | 8 | 9 | * | * | * | * | * | * |
| * | * | * | 1 | 2 | 3 | * | * | * |
| * | * | * | 4 | 5 | 6 | * | * | * |
| * | * | * | 7 | 8 | 9 | * | * | * |
| * | * | * | * | * | * | 1 | 2 | 3 |
| * | * | * | * | * | * | 4 | 5 | 6 |
| * | * | * | * | * | * | 7 | 8 | 9 |

Fig. 6

| 1 | 2 | 3 | 4 | 5 | 6 | 7 | 8 | 9 |
|---|---|---|---|---|---|---|---|---|
| 4 | 5 | 6 | 7 | 8 | 9 | 1 | 2 | 3 |
| 7 | 8 | 9 | 1 | 2 | 3 | 4 | 5 | 6 |
| 2 | 3 | 1 | * | * | * | * | * | * |
| 5 | 6 | 4 | * | * | * | * | * | * |
| 8 | 9 | 7 | * | * | * | * | * | * |
| 3 | 1 | 2 | * | * | * | * | * | * |
| 6 | 4 | 5 | * | * | * | * | * | * |
| 9 | 7 | 8 | * | * | * | * | * | * |

Fig. 7

| 1 | 2 | 3 | 4 | * | 6 | 7 | 8 | 9 |
|---|---|---|---|---|---|---|---|---|
| 4 | 5 | 6 | * | * | * | 1 | 2 | 3 |
| 7 | 8 | * | * | * | * | * | 5 | 6 |
| 2 | * | * | * | * | * | * | * | 1 |
| * | * | * | * | * | * | * | * | * |
| 8 | * | * | * | * | * | * | * | 7 |
| 3 | 4 | * | * | * | * | * | 1 | 2 |
| 6 | 7 | 8 | * | * | * | 3 | 4 | 5 |
| 9 | 1 | 2 | 3 | * | 5 | 6 | 7 | 8 |

Fig. 8

| 1 | 2 | 3 | 4 | 5 | 6 | 7 | 8 | 9 |
|---|---|---|---|---|---|---|---|---|
| 4 | 5 | 6 | 7 | 8 | 9 | 1 | 2 | 3 |
| 7 | 8 | 9 | 1 | 2 | 3 | 4 | 5 | 6 |
| 2 | 3 | 4 | 5 | 6 | 7 | 8 | 9 | 1 |
| 5 | 6 | 7 | 8 | 9 | 1 | 2 | 3 | 4 |
| 8 | 9 | 1 | 2 | 3 | 4 | 5 | 6 | 7 |
| * | * | * | * | * | * | * | * | * |
| * | * | * | * | * | * | * | * | * |
| * | * | * | * | * | * | * | * | * |

Fig. 9

| 1 | 2 | 3 | 4 | 5 | 6 | 7 | 8 | 9 |
|---|---|---|---|---|---|---|---|---|
| 4 | 5 | 6 | 7 | 8 | 9 | 1 | 2 | 3 |
| 7 | 8 | 9 | 1 | 2 | 3 | 4 | 5 | 6 |
| 2 | 3 | 4 | 5 | 6 | 7 | * | * | * |
| 5 | 6 | 7 | 8 | 9 | 1 | * | * | * |
| 8 | 9 | 1 | 2 | 3 | 4 | * | * | * |
| 3 | 4 | 5 | * | * | * | * | * | * |
| 6 | 7 | 8 | * | * | * | * | * | * |
| 9 | 1 | 2 | * | * | * | * | * | * |

Fig. 10

| 1 | * | * | 7 | * | * | 4 | * | * |
|---|---|---|---|---|---|---|---|---|
| * | 2 | * | * | 8 | * | * | 5 | * |
| * | * | 3 | * | * | 9 | * | * | 6 |
| 7 | * | * | 4 | * | * | 1 | * | * |
| * | 8 | * | * | 5 | * | * | 2 | * |
| * | * | 9 | * | * | 6 | * | * | 3 |
| 4 | * | * | 1 | * | * | 7 | * | * |
| * | 5 | * | * | 2 | * | * | 8 | * |
| * | * | 6 | * | * | 3 | * | * | 9 |

Fig. 11